\documentclass[twocolumn,showpacs,showkeys,preprintnumbers,amsmath,amssymb]{revtex4}

\usepackage{graphicx}
\usepackage{dcolumn}
\usepackage{bm}
\usepackage{subfigure}
\newcommand{\ave}[1]{\langle #1 \rangle}

\begin{document}

\preprint{}

\title{Disaster Management in Scale-Free Networks:
Recovery from and Protection Against Intentional Attacks}%

\author{Behnam A. Rezaei}
\email{behnam@ee.ucla.edu}
\affiliation{Department of Electrical Engineering \\ University of California, Los Angeles}
\author{Nima Sarshar}
\email{nima@ee.ucla.edu}
\affiliation{Department of Electrical Engineering \\ University of California, Los Angeles}
\author{P. Oscar Boykin}
\email{boykin@ece.ufl.edu}
\affiliation{Department of Electrical and Computer Engineering \\ University of Florida, Gainesville}
\author{Vwani P. Roychowdhury}
\email{vwani@ee.ucla.edu}
\affiliation{Department of Electrical Engineering \\ University of California, Los Angeles}
%


\begin{abstract}
Susceptibility of scale free Power Law (PL) networks to attacks
has been traditionally studied in the context of what may be
termed as {\em instantaneous attacks}, where  a randomly selected
set of nodes and edges are deleted while the network is kept {\em
static}. In this paper, we shift the focus to the study of {\em
progressive} and instantaneous attacks on {\em reactive} grown and
random PL networks, which can respond to attacks and take remedial
steps. In the process, we present several techniques that managed
networks can adopt to minimize the damages during attacks, and
also to efficiently recover from the aftermath of successful
attacks.   For example, we present (i) compensatory dynamics that
minimize the damages inflicted by targeted progressive attacks,
such as linear-preferential deletions of nodes in grown PL
networks; the resulting dynamic naturally leads to the emergence
of networks with PL degree distributions with exponential cutoffs;
(ii) distributed healing algorithms that can scale the maximum
degree of nodes in a PL network using only local decisions, and
(iii) efficient means of creating giant connected components in a
PL network that has been fragmented by attacks on a large number
of high-degree nodes. Such targeted attacks are considered to be a
major vulnerability of PL networks; however, our results show that
the introduction of  only a small number of random edges, through
a {\em reverse percolation} process, can restore connectivity,
which in turn allows restoration of other topological properties
of the original network. Thus, the scale-free nature of the
networks can itself be effectively utilized for protection and
recovery purposes.

\end{abstract}

\maketitle

\section{Introduction}\label{sec:intro}
A large number of networks in different contexts have been found
to have scale free structures( see \cite{albe02} and references
there in) characterized by Power-Law degree distributions. For a
power-law (PL) degree distribution, the probability of a randomly
chosen node to have degree $k$ scales as $P(k)\propto k^{-\gamma}$
for large $k$; $\gamma$ is referred to as the exponent of the
distribution. Moreover, a PL distribution is  considered to be
{\em heavy tailed} if $2< \gamma\leq 3$, i.e., when the mean is
bounded but the variance is unbounded. While many of these
networks have evolved naturally, driven by dynamical processes
over which we do not have much control, there are several classes
of other networks, where the underlying dynamics can be altered to
make sure that the resulting networks not only have scale-free
structures, but are also resilient to external attacks. Examples
of such designer complex networks include, the Internet, national
power grids and other infrastructure related networks, computer
and communication networks, and more recently Peer-To-Peer (P2P)
networks. With the economy going more global every year, the
emergence of different kinds of complex networks that interconnect
distributed centers of communication, finance, and manufacturing,
will only see a rapid growth. In such networks,  attacks are a
fact of life, and simple attacks, such as Denial-of-Service (DOS),
can cripple hubs and other nodes, leading to   severe disruptions
of services. Understanding the effect of attacks, and mechanisms
to respond to attacks is thus of great practical importance to
many network based systems.

As reviewed in the following, the study of attacks, including both targeted and random, in scale-free PL networks have been mostly restricted to the case of instantaneous or massive attacks carried out on passive networks. In this paper, we shift the focus to the study of progressive and rapid attacks on networks that can respond in an active fashion. In the process, we design several techniques that managed networks can adopt to minimize the damages, and also to efficiently recover from the aftermath of successful attacks.  We find that the {\em scale-free nature of the networks can be judiciously utilized for such protection and recovery purposes}.

\subsection{{Instantaneous} vs. {Progressive} Attacks and {Reactive} Vs. {Non-Reactive} Networks}
Susceptibility of networks to attacks has been traditionally
studied in the context of what may be termed as {\em instantaneous
attacks}, where  a randomly selected set of nodes and edges are
deleted while the network is kept static. Hence, for all purposes,
the targeted nodes and edges are deleted simultaneously.  The
effect of such instantaneous attacks is then studied in terms of
the {\em connectivity structure} of the network that is left
behind after the attack, i.e., how many connected components are
there in the compromised network, and if there exist {\em giant
connected components} that contain a constant fraction of the
remaining nodes.  It is well known, for example, that in random
Erd\"{o}s-R\'{e}nyi (ER) networks\cite{Bollobas01} it is
sufficient to remove nodes independently with a probability  {\em
above a certain threshold} in order to break up the network into
many subnetworks that are {\em all small in size}. Thus, random ER
networks are considered susceptible to random deletions of nodes,
or just {\em instantaneous random  attacks} (\underline{\em
IRAs}). The case where the
deleted nodes are picked non-uniformly randomly (e.g., {\em
preferentially} with respect to their degrees) has also been
studied; for the purposes of this paper, such attacks will be
termed as {\em Instantaneous Targeted  Attacks} (\underline{\em
ITAs}).

From a network management perspective, a key feature of such attack models is that the network remains {\em passive} or {\em non-reactive} during an attack: the removal of nodes and edges occurs without allowing the network to  take actions  (for example, inserting extra edges or allowing a deleted node to rejoin the network) to minimize the disruptive effects of the attacks. We shall refer to such networks as \underline{\em non-reactive} networks. On the other hand, if a networked system  takes active compensatory measures to maintain its integrity during an attack, or if it takes measures to recover from the damages inflicted by an instantaneous attack, then we will refer to such networks as \underline{\em reactive} networks.  Moreover, an attack that takes place at rates comparable to the response time of a network will be referred to as a \underline{\em progressive} or {\em gradual} attack (as opposed to instantaneous attacks).  In this paper, instead of only studying the case of  instantaneous attacks on non-reactive networks, {\em we explore  the cases of both progressive and instantaneous attacks on reactive and dynamic networks}, and study how the underlying networks can protect against, and recover from such attacks.

\subsection{Instantaneous Attacks on  {Random} and  {Grown} Scale-Free PL Networks}
Both IRAs and ITAs have been studied using the concept of
percolation theory.
Consider a network of size $N$ in which the largest connected
component comprises a fraction  $\lambda$ of the nodes. The
site percolation process proceeds as follows: Take a
constant probability $0\leq p\leq 1$ (i.e., independent of the size parameter $N$) called the {\em percolation probability}.
Delete each node in the network independently with probability $1-p$ and retain
it with probability $p$. For large $N$, the resulting network will
have almost $pN$ nodes. Out of these $pN$ nodes, a set of $S(N,p)$ nodes  will form a single connected component of the largest size. The main lesson of percolation theory is that
for many families of graphs there exists a $p_c >0$, such that if
$p>p_c$ then  $\displaystyle \lim_{N\rightarrow
\infty}\frac{S(N,p)}{Np}= \lambda'(p)> 0$, and if  $p<p_c$, then
$\displaystyle \lim_{N\rightarrow
\infty}\frac{S(N,p)}{Np}= 0$. The critical percolation probability, $p_c$, is called the  {\em percolation threshold.}  This \emph{uniform instant
site percolation process} is a particular case of IRAs introduced earlier.

One can generalize the above-mentioned percolation concept of picking nodes randomly but uniformly, to where the nodes to be deleted are picked randomly, but with a distribution based on their degrees. For example, the case of {\em targeted instant site percolation process} (or ITA's) might consist of {\bf (i)} deleting all  nodes of degree greater than a pre-specified value of say $k_0$, or {\bf (ii)}  if  the fraction of nodes  with degree $k$ in the original network is $p_k$, then the fraction  of nodes of degree $k$ after the attack is  reduced to $p_{k}^\prime = bk^{-q}p_k$, where $b$ is a normalization constant. In both cases, the high-degree nodes are the targets of severe attacks, and the low degree nodes are mostly left alone. 

The percolation properties of \emph{random heavy-tailed
scale-free} Power-Law networks have been studied
extensively\cite{CNSW00,AJB00,BKMPRSTW00} and provide a mixed
message when it comes to their vulnerabilities to IRAs and ITAs.
For IRAs, the percolation theory reveals a very promising fact:
{\em The percolation threshold of these graphs is zero}. That is,
no matter how small the percolation probability, $p$, is, the
remaining nodes in the percolated network has a giant connected
component. Thus, random PL networks can withstand  IRAs with
arbitrarily high rates. Previous studies, however, have also shown
that the {\em random PL networks are more vulnerable to ITAs},
e.g., removal of a large number of only high degree nodes from
scale-free networks is enough to fragment the network such that a
giant connected component does not exist\cite{CEBH01,DM01,CNSW00}.
Thus, by removing a constant fraction of all nodes
preferentially, one can destroy the connectivity of the network;
note that {\em the attacker still has to remove almost all the
high-degree nodes} to do so, which might be a difficult task to
accomplish \cite{CEBH01}.

In many situations, however, {\em the scale-free PL networks are
not truly random}, but can be modelled as being grown by dynamical
rules, where nodes and edges are added at certain rates. Such
networks will be referred to in this paper as {\em grown
scale-free networks}.  For example, the simple preferential
attachment dynamic and its variants can give rise to scale-free
grown networks with several tunable topological characteristics.
 Unlike the case of the random scale-free networks, currently
there is {\em no instantaneous percolation theory for these grown
scale-free networks}. Nevertheless, empirical studies suggest that
these networks are also both resilient to random deletions of
their nodes (IRAs), and vulnerable to targeted attacks just as in
the case of random scale-free networks \cite{PKPLGU03}.

For both random and grown PL networks, while we know that they are
vulnerable to severe ITAs, the issue of how to recover from such attacks
and glue the fragmented network back together efficiently has not
been addressed. We show in Section~\ref{sec:glueing} how the PL
structure of the network can in fact be an asset in this recovery
process.

\subsection{Progressive Attacks and {Grown} Scale-Free Networks}
Recall that in progressive or gradual attacks, the deletions of nodes and edges take place at rates comparable to those at which the dynamics of the grown network itself operates at. For example, a {\em progressive attack} might correspond to a scenario, where randomly chosen existing nodes in the network (picked preferentially or uniformly with respect to the degree of a node) are removed at the {\em same rate} at which  {\em the new nodes join in}. What do such progressive attacks do to the grown networks as opposed to the well studied case of instantaneous attacks?

Consider the simple case of linearly preferentially grown network
\cite{BA}, where in addition to a node joining the network at each
time step, a {\em uniformly} randomly chosen existing node in the
network is deleted with probability $c$ at each step. Such a
dynamic is considered in detail in \cite{uss,Dog}. It turns out
that unlike the instantaneous attack case, where a grown network
is resistant to IRAs, {\em a grown network} is very {\em
vulnerable to random progressive attacks}. However, the {\bf
connectivity structure of the attacked network is no longer a
relevant measure} to study the effect or severity of the
progressive attacks;  the network almost always remains connected,
or has a giant connected component. As shown in \cite{SR04}, {\em
the damage to the network manifests itself by forcing the network
to rapidly lose its heavy-tailed distribution} (i.e., the PL
exponent becomes much grater than 3, even as $c$ increases only
marginally), and the resulting grown network starts resembling
networks with exponential degree distributions under the attack.

A {\em reactive} grown network, however, may take {\em remedial
actions}, and one might ask if there exist compensatory dynamics
that will restore the heavy-tailed distribution even in the
presence of the random progressive attack. It was shown in
\cite{SR04} that indeed the linear-preferential attachment
dynamics can be modified in a very simple and local fashion to
preserve the heavy-tailed degree distribution and the scale-free
nature of the grown network.  The compensatory procedure is
intuitive and greedy:  {\em whenever a node loses a connection},
which can only happen when a neighboring node is deleted due to
the progressive attack, {\em it compensates for it by making a new
preferential connection} with a certain probability $n$. Thus, the
{\em dynamics remain strictly local} (each node reacts only if it
is directly impacted by the attack dynamic) and yet, the end
result is that the damages to the global topological properties of
the network are repaired, even in the limit of extremely high
deletion rates (i.e., $c\rightarrow 1$).

This particular case study brings out two important differences
between instantaneous and progressive attacks: (i) While a grown
network might be resistant to IRAs, it can be extremely vulnerable
to progressive random attacks. The damage to the network is no
longer in terms of a loss in  connectivity, but rather in terms of
other topological properties of the network.  (ii) It is, however,
possible to make the dynamical rules of the grown networks to be
reactive, and generate networks that are extremely resistant to
both gradual and instantaneous random failures and attacks.

\subsection{Protecting Against and Recovering from Attacks in Reactive Scale-Free Networks: A Summary of Results}
We first consider {\em protection against progressive or gradual
targeted attacks in grown networks.} The need to address this type of
intentional attacks is {\em particularly urgent in designer complex
networks, such as the P2P networks}, where performing a
comprehensive large-scale attack on all the high-degree nodes is
an expensive, and often, a very difficult task. However, {\em
gradual deletions of high-degree nodes} by first crawling the
network and identifying the high-degree nodes that serve as
conduits for communication among low-degree nodes, and then
attacking some of these nodes {\em might be quite feasible}.

Clearly, if no precaution or compensatory action is taken against
such an attack,  then as exemplified in the case of progressive
random attacks\cite{SR04}, one would lose key topological
features, including a loss of its heavy tailed degree
distribution.  How to equip the network with proper feed-back
strategies to mitigate the damaging effects of the attacks and
failures? These feedback strategies, moreover, must obey some
stringent criteria: They must be local, in the sense that they
should be triggered and adjusted only based on first neighbor
information. In Section~\ref{sec:grow} we introduce one such
dynamical compensatory algorithm to mitigate the effect of
linear-preferential attacks. In particular we show that, only with simple
local modifications to the dynamics, the network can restore much
(but not all) of its heavy-tail. {\em The resulting degree
distribution is shown to be a power-law with an exponential cutoff
at a point that is inversely proportional to the rate of the
targeted attack}. Again, there is always a  giant connected
component, and the main effect of the attack  is to introduce an
exponential cutoff, and lower the PL exponent marginally.  Thus,
while the attacked network does lose its unbounded variance, the
compensatory dynamics manage to preserve the overall PL degree
distribution, and as shown next, keep the network in a state, from
which it can recover in a local fashion.

We next show in Section~\ref{subsec:healing} that a {\em
preferentially attacked  network can 
perform large-scale network repairing and maintenance operations}
and selectively add edges, so as to {\em increase the $k_{max}$ by
any desired scaling factor}; thus, this can repair the cutoff problem
resulting from linear-preferential attacks. The recovery procedure is local, in
the sense that each node independently decides how many
preferential connections to create, and no global coordination is
necessary. This procedure retains the exponent of the PL
distribution, and only increases the maximum-degree of the
distribution, so that the exponential cut-off point is not a
limiting factor. 

Next, we consider the case of {\em recovery from a large-scale
targeted attack}. A heavy enough and instantaneous targeted
attack will finally fragment any static or dynamic network. In
Section~\ref{sec:glueing} we for the first time consider the
challenging problem of repairing a complex network fragmented by
targeted attacks. The disaster recovery consists of two distinct
phases, and the first  is to {\em recover the lost connectivity}. We show that
with only a few essential communication paths one can create a
giant connected component and glue the network fragments. In
particular, we first show (both analytically and numerically) that
when scale-free networks are fragmented due to targeted attacks,
it results in small-size connected components, the {\em size
distribution} of which is again {\em heavy-tailed}. This allows us to
prove that the connectivity of the whole fragmented network can be
restored with only a few successful random connections, via the
process of {\em reverse percolation}. Thus, the nodes of the
network will be able to communicate to each other to transfer
\emph{vital} "low-rate" messages, as long as a few random connections among the nodes can be established. {\em Recovery of the
topology,} can then be achieved, once connectivity is established. Such a recovery step  could consist of applying one of the many dynamical rules \cite{SR04,albe02,BA} that would allow it to
regain its scale-free structure.

\section{Reactive Grown Networks in Presence of Linear-Preferential Attacks and Compensation}
\label{sec:grow} 
\begin{figure}[h]
\centering \scalebox{0.8}{} \centering
     \scalebox{0.4}{\includegraphics{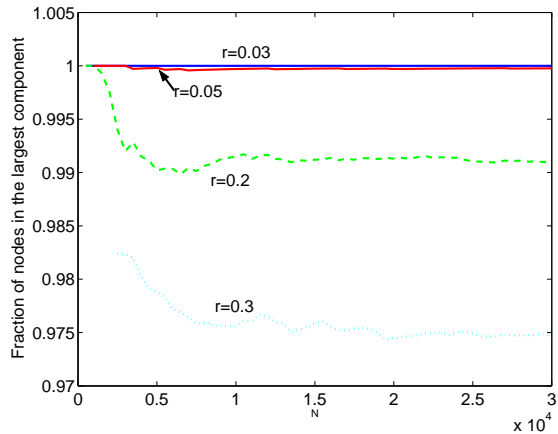}}
    \caption{The fraction of nodes in the largest connected component for various rates of targeted attack when $m=4$. Even for rates as high as 30\%, more than 97\% of the nodes belong to a single component. }
    \label{fig:simcomp}
\end{figure}

This section considers a progressive attack,
where at each time step in addition to a node joining the network,
a preferentially chosen node is deleted with probability $r$.
Nodes that loose neighbors to attack don't sit still, instead they
react and replace those lost neighbors; moreover the attacked
nodes rejoin the network as new nodes. We show that such a
compensatory dynamic in the presence of linear-preferential
progressive attacks naturally leads to a {\em PL degree
distribution with an exponential cutoff}, where the cut-off
depends on the preferential deletion rate. Thus, while the attack
bounds the maximum degree of the distribution, the compensatory
protocol is able to preserve the exponent of the PL distribution.
Moreover, as shown in our simulations, as long as each incoming
node makes $m\geq 2$ random preferential connections, there always
exists a giant connected component, even for very high rates of
preferential attacks (see Fig. \ref{fig:simcomp} for an example);
thus the loss of connectivity is not one of the damaging effects
of such progressive attacks.

Our model is a time dynamic one.  At each step:
\begin{enumerate}
    \item A new node $s$ is added to the network and makes $m$
    preferential attachments.
    \item With probability $r$ a preferentially selected node $w$ is chosen. The preferential selection procedure selects the node $w$ with degree $k(w,t)$ with probability
    $k(w,t)/\sum_i{k(i,t)}$.
    The selected node $w$ is then deleted from the network, were deletion process is:
    \begin{itemize}
        \item Delete node $w$ and all its edges. Then $w$ starts
        as a new node and makes $m$ preferential attachments.
        \item For all nodes $z$ that were connected to $w$ and have lost
        an edge, each one compensates by adding an edge
        preferentially.
    \end{itemize}

\end{enumerate}

\subsection{Analysis}\label{subsec:analysis}
We adopt the same rate equation approach as \cite{Dog1,SR04} for
our analysis. Label each node with its insertion time to network,
$i$. Define degree of $i^{th}$ node at time $t$ as $k(i,t)$. When
a node is deleted its joining time is reset. Let
$S(t)=\sum_i{k(i,t)}=2E(t)= 2(m+r)t$ and total number of nodes at
time $t$, $N(t)=t$. The mean degree is $\ave{k}=2(m+r)$. We want
to find $P(k)$, i.e. probability that a randomly chosen node has
degree $k$ at steady state ($t\rightarrow \infty$). We also define
$f(t)$ as average number of edges deleted at time $t$ when a node
is deleted preferentially, that is the average degree of a node
chosen preferentially: $f(t)=\frac{\ave{k(i,t)^2}}{\ave{k}}$. Note
that obviously $\ave{k}\le f(t) \le E(t)$. Define the probability
that $i^{th}$ node is not deleted before time t (is still in the
network) as $D(i,t)$. The initial conditions are $D(i,i)=1$, and
$k(i,i)=m$. Next we write master equations for $k(i,t)$ and
$D(i,t)$

\begin{equation}
    \frac{\partial k(i,t)}{\partial
    t}=m\frac{k(i,t)}{S(t)}+rm\frac{k(i,t)}{S(t)}+rf(t)\frac{k(i,t)}{S(t)}
\label{masterEQ}
\end{equation}
where the first term corresponds to $m$ preferential attachments,
the second term corresponds to deleted node rejoining the network,
and the last term represents preferential compensation of on
average $rf(t)$ edges. Note that since each node compensates for
lost edges there is no negative term in the equation. Also at time
$t+1$ the probability that $i^{th}$ node still exist in the
network is given by:
\begin{eqnarray}
    D(i,t+1)&=&D(i,t)\left(1-r\frac {k(i,t)}{S(t)}\right)\nonumber\\
    D(i,t+1) - D(i,t) &=& -r D(i,t)\frac{k(i,t)}{S(t)}\nonumber\\
    \frac{\partial D(i,t)}{\partial t}
    &=&-r D(i,t)\frac{k(i,t)}{S(t)}\nonumber\\
    \ln \frac{D(i,s)}{D(i,i)}&=&-\int_i^s{\frac{r k(i,t)}{2(m+r)t}dt}\nonumber\\
    &=& -\frac{r}{\ave{k}}\int_i^s{\frac{k(i,t)}{t}dt}\nonumber\\
    D(i,s)&=&\exp\left(
    -\frac{r}{\ave{k}}\int_i^s{\frac{k(i,t)}{t}dt}\right)
    \label{eq:d_of_k}
\end{eqnarray}

Define $\widetilde{k(i,s)}=\int_i^s \frac{k(i,t)}{t}dt$, then
$D(i,s) = \exp(-r\widetilde{k(i,s)}/\ave{k})$.  Note if $k(i,t)$
is lower bounded by  $C t^\beta$ then
$\widetilde{k(i,t)}\ge\frac{1}{\beta}k(i,t)$:
\begin{eqnarray}
k(i,t)&\ge&C t^{\beta}\nonumber\\
\widetilde{k(i,t)}&\ge&
\frac{k(i,t)}{\beta}\label{eq:tildek_bound}
\end{eqnarray}

In order to solve for $k(i,t)$, we need to know $f(t)$, which
depends on $k(i,t)$:
\begin{equation}
\label{eq:f_sum}
f(t)=\frac{\ave{k^2}}{\ave{k}}=\frac{1}{N(t)\ave{k}}\sum_{i=0}^t
k(i,t)^2 D(i,t)
\end{equation}
When Equation \ref{eq:tildek_bound} is valid then:
\begin{eqnarray*}
f(t)&=&\frac{1}{N(t)\ave{k}}\sum_{i=0}^t k(i,t)^2 D(i,t)\\
    &=&\frac{1}{t\ave{k}}\sum_{i=0}^t k(i,t)^2 \exp\left(-\frac{r}{\ave{k}}
                                               \widetilde{k(i,t)}\right)\\
    &\le& \frac{1}{t\ave{k}}\sum_{i=0}^t k(i,t)^2 \exp\left(-\frac{r}{\beta\ave{k}}
                                               k(i,t)\right)
\end{eqnarray*}
Since $x^2 \exp(-\alpha x)\le\frac{4e^{-2}}{\alpha^2}$:
\begin{eqnarray}
f(t) &\le& \frac{1}{t\ave{k}}\sum_{i=0}^t \frac{(\beta\ave{k})^2}{r^2} 4e^{-2}\nonumber\\
     &=& \frac{4e^{-2}\beta^2 \ave{k}}{r^2}\label{eq:const_f}
\end{eqnarray}
Thus we see that if $k(i,t)$ is a finite polynomial of positive
powers of $t$, then $f(t)$ is a constant.

We know that $f(t)\le E(t) = (m+r)t$, but edge deletion may reduce
it further.  Hence we consider the case where $f(t)=at^b$. We
consider this for two cases, first where $b=0$, and second where
$b>0$. If $f(t)=a$ then Eqn. \ref{masterEQ} becomes
\begin{eqnarray*}
    \frac{\partial k(i,t)}{\partial t}&=&
           \left(m(1+r)+ra\right)\frac{k(i,t)}{S(t)}\\
       &=&\left(m+r + r(m+a-1)\right)\frac{k(i,t)}{2(m+r)t}\\
       &=&\left(\frac{1}{2}+\frac{r(m+a-1)}{2(m+r)}\right)\frac{k(i,t)}{t}
\end{eqnarray*}
so, using $\epsilon\equiv \frac{r(m+a-1)}{2(m+r)}$
\begin{eqnarray}
  \frac{\partial k(i,t)}{k(i,t)}&=&
    \left(\frac{1}{2} + \epsilon\right)1/t\nonumber\\
  \ln\frac{k(i,t)}{k(i,i)}&=&\left(\frac{1}{2}+\epsilon\right)\ln(t/i)\nonumber\\
    k(i,t)&=& m\left(\frac{t}{i}\right)^{\frac{1}{2}+\epsilon} \label{eq:k_of_t}
\end{eqnarray}
Eqn. \ref{eq:const_f} already showed that if $k(i,t)$ grows as a
power of $t$, $f(t)$ is constant. The above shows that when we
assume that $f(t)=a$, we see that $k(i,t)$ grows as a power of
$t/i$.

We calculate $P(k,t)$ from $k(i,t)$ and $D(i,t)$ using Eqn.
\ref{eq:d_of_k}:
\begin{eqnarray*}
    D(i,s)&=&\exp\left(
    -\frac{r}{\ave{k}}\int_i^s{\frac{k(i,t)}{t}dt}\right) \\
    &=&\exp\left(
    -\frac{r}{\ave{k}}
    \int_i^s{ m\left(\frac{t}{i}\right)^{1/2+\epsilon}\frac{1}{t}dt}\right)\\
    &=&\exp\left(
    -\frac{rm}{i^{\frac{1}{2}+\epsilon}\ave{k}}
    \int_i^s{ t^{\epsilon-\frac{1}{2}}}dt\right)\\
    &=&\exp\left(-\frac{r}{(\frac{1}{2}+\epsilon)\ave{k}}(k(i,t)-m)\right)
\end{eqnarray*}
\begin{eqnarray}
    P(k,t)&=&\frac 1 {N(t)}. \Sigma_{i:k(i,t)=k}{D(i,t)}\nonumber\\
    &=&\frac 1 t D(i,t) |\frac{\partial i}{\partial k}|
\end{eqnarray}
If $k(i,t)=m(t/i)^{1/2+\epsilon}$:
\begin{eqnarray*}
k(i,t)&=&m(t/i)^{\frac{1}{2}+\epsilon}\\
(k(i,t)/m)^{\frac{2}{1+2\epsilon}} &=& t/i\\
i &=& t \left(\frac{k}{m}\right)^{\frac{-2}{1+2\epsilon}}\\
\frac{\partial i}{\partial k} &=& t\frac{2}{1+2\epsilon}
                         \left(\frac{k}{m}\right)^{\frac{-3 - 2\epsilon}{1+2\epsilon}}
\end{eqnarray*}
Thus:
\begin{eqnarray*}
P(k,t)&=& \frac{1}{t}D(i,t)|\frac{\partial i}{\partial k}|\\
      &=& \frac{1}{t}\exp\left(-\frac{r}{(\frac{1}{2}+\epsilon)\ave{k}}(k-m)\right)
          t\frac{2}{1+2\epsilon}
                         \left(\frac{k}{m}\right)^{\frac{-3 - 2\epsilon}{1+2\epsilon}}\\
      &=& \exp\left(-\frac{r}{(\frac{1}{2}+\epsilon)\ave{k}}(k-m)\right)
          \frac{2}{1+2\epsilon}
                         \left(\frac{k}{m}\right)^{\frac{-3 - 2\epsilon}{1+2\epsilon}}
\end{eqnarray*}
So, when we assume that $f(t)=a$, or that $\ave{k^2}=a \ave{k}$,
we see that we get a power-law degree distribution with an
exponential cut-off at $k_c =
\ave{k}(\frac{1}{2}+\epsilon)/r=(m/r+1)(1+2\epsilon)$.  As
$r\rightarrow 0$, $k_c\rightarrow \infty$, as expected.

Now we consider the case of $f(t)=at^b$ with $b>0$:
\begin{eqnarray*}
    \frac{\partial k(i,t)}{\partial t}&=&
           \left(m(1+r)+rat^b\right)\frac{k(i,t)}{S(t)}\\
           &=&\left(\frac{m+r+r(m-1)}{2(m+r)}\frac{1}{t} +
       \frac{ra}{2(m+r)}t^{b-1}\right)k(i,t)\\
  \ln\frac{k(i,t)}{k(i,i)}&=&\left(\frac{m+r+r(m-1)}{2(m+r)}\ln t +
       \frac{ra}{2(m+r)}\frac{t^{b}}{b}\right)\\
\ln\frac{k(i,t)}{m} &=&\left(\frac{1}{2} + \beta\right)\ln (t/i) +
         \gamma\frac{t^b-i^b}{b}\\
k(i,t)&=& m \left(\frac{t}{i}\right)^{\frac{1}{2}+\beta}
\exp\left(\frac{\gamma}{b}(t^b-i^b)\right)\\
      &\ge& m \left(\frac{t}{i}\right)^{\frac{1}{2}+\beta}
\end{eqnarray*}
with $\beta=\frac{r(m-1)}{2(m+r)}$ and $\gamma=\frac{ra}{2(m+r)}$.
We see that even if we assume that $f(t)=at^b$, then we find that
$k(i,t)\ge m(t/i)^{1/2+\beta}$, but then according to Eqn.
\ref{eq:const_f}, $f(t)$ is upper bounded by a constant, which
implies that $k(i,t)$ only grows as a power of $t$ as we saw in
Eqn. \ref{eq:k_of_t}.

This result clearly shows that preferential deletion of nodes will
impose an exponential cut-off on power-law degree distribution of
the growing networks thus removing high degree tail of the network
which are essential for its functionality \cite{SBR04}.

\subsection{Simulations} \label{sec:simulations} 
\begin{figure}[h]
\centering \scalebox{0.8}{} \centering
     \scalebox{0.5}{\includegraphics{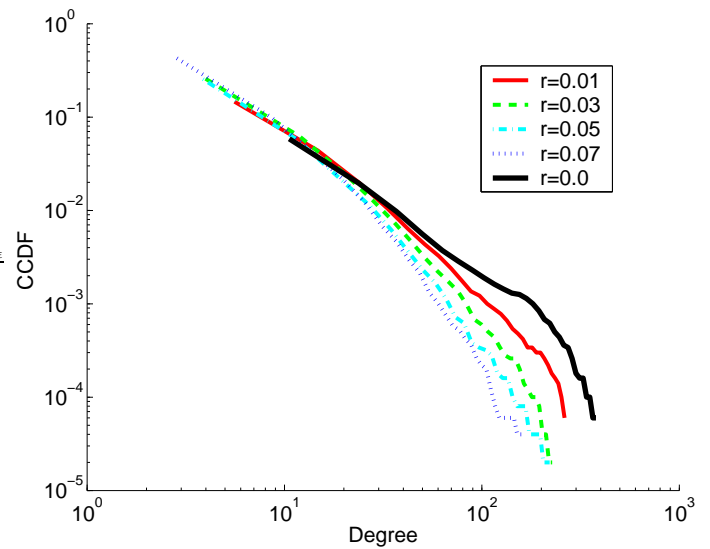}}
    \caption{The CCDF of the degree distribution for various preferential node deletion probabilities for a network of size 50,000 and $m=2$.}
    \label{fig:prefdelete}
\end{figure}

We have performed extensive
Monte Carlo simulations. We grow the network where at each step a
node is added and makes $m=2$ preferential attachments. Then with
probability $r$ we delete a node, choosing a node to be deleted
preferentially. Fig. \ref{fig:prefdelete} shows the effect of low
rate preferential deletion as imposing an exponential cutoff on
the scale-free degree distribution.

\section{Decentralized and Distributed Healing of Sharp Cutoffs in PL Networks }\label{sec:healing}
We consider a connected PL network where the tail (i.e., the set
of high degree nodes) is removed.  We first show how a local
compensation process, involving the creation of new links, can
restore the maximum degree to a multiple of the existing cutoff.
Each node decides randomly and independently how many preferential
edges to insert, and thus the healing process does not need any
central coordination mechanism. Such a healing process can be
periodically applied by, for example, a network created by the
dynamics described in the previous section, where a cutoff in the
degree distribution is naturally introduced.

\subsection{Healing Process}\label{subsec:healing}
Consider a \emph{short tailed} power-law network of exponent
$\alpha$ and maximum degree $k_0$. The objective of the healing
process is to increase the maximum degree of the distribution by a
stretch-factor $w = k_{max}/k_0$, so that the maximum degree after
the healing process will increase from $k_0$ to $k_{max}$, while
the power-law exponent remains the same.

\textbf{Healing Algorithm} Given a PL network with exponent
$\alpha$, each node $i$ independently decides to compensate with
some probability $p$. The compensation process involves making
$(w-1)k_i$ new \textit{preferential} connections where $k_i$ is
the degree of the $i^{th}$ node.

Let $P(k)$ and $P'(k)$ be the degree distributions before and
after the healing process respectively, then one can easily show
that given $P(k)$ is a short tailed power-law with maximum degree
$k_0$ and $p\approx w^{-\alpha}$, $P'(k)$ follows a power law
distribution as $\lambda{(w,\alpha)}k^{-\alpha}$. Where $\lambda$
is a constant depending on $w$ and $\alpha$ and cut off is
increased to $k_{max}$. In other words, the effect of the healing
process is simply to stretch the degree distribution by a linear
factor. Now since the degree distribution is
\emph{\textbf{scale-free}} to begin with, such stretching will not
change the distribution. Thus power-law degree distribution of the
original network is the key for the simple healing process to
succeed.

\subsection{Autonomous Healing}\label{subsec:distselfhealing}
So far we have considered a {\em static case}, i.e., we are given a
network with a sharp cutoff, and the nodes randomly decide to introduce new edges to
restore the cutoff to a desired value, $k_{max}$, while retaining the same
PL exponent, $\alpha$, as before. One can modify this static scenario
to an {\em adaptive version}, where instead of all the nodes acting at once,
each node reacts whenever it loses an edge unannounced, i.e., due to an attacked
node going down.  The idea is to naturally detect a heavy attack and initiate the
healing process. \\
\textit{Feedback Algorithm}: Each node $i$ when losing an edge
{\em without prior notice} performs the  healing algorithm of
Section \ref{subsec:healing} with probability $1/k_i$, where $k_i$
is the degree of the $i^{th}$ node.

Clearly any targeted deletion of nodes in the tail of the degree
distribution, i.e., high degree nodes, will result in the deletion
of a constant fraction of the edges of the system. If we further
neglect the degree correlation of the nodes, the probability of
any edge being deleted when nodes of degree greater than $k_0$ are
deleted will be given by:
\begin{equation}
\widetilde{p} \approx \frac{\Sigma_{k_0}^{k_{max}}{kP(k)}}{E}=
\lambda k_0^{2-\alpha}\ .
\end{equation}
for some constant $\lambda$ in the order of 1. Then, the
probability of a node with degree $k$ losing an edge is
$\widehat{P}_k \approx k\widetilde{p}$. Thus the probability of a
node initiating the healing process is approximately
$\widehat{P}_k/k=\widetilde{p}$, a constant depending on only the
intensity of the attack; We must re-state that we have not
considered second order effects and degree correlations here. This
algorithm is basically intended to detect any large scale
instantaneous deletion of a fraction of network edges. Thus, {\em
the algorithm ensures that a constant fraction of the nodes will
always perform the healing algorithm in the case of a large enough
attack. }

\begin{figure}%
  \centering
     \subfigure[]
{
        \centering
         \scalebox{0.5}{\includegraphics{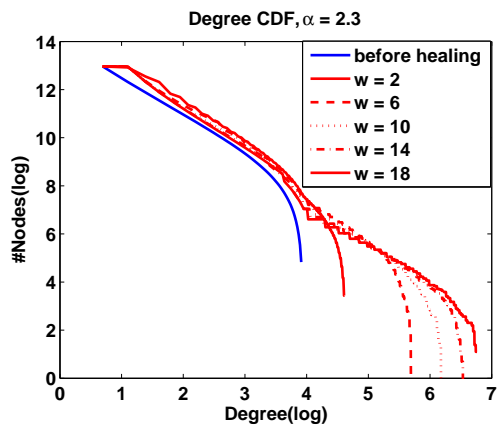}}}
     \hspace{0.2in}%
    \subfigure[]
       {%
        \centering
         \scalebox{0.5}{\includegraphics{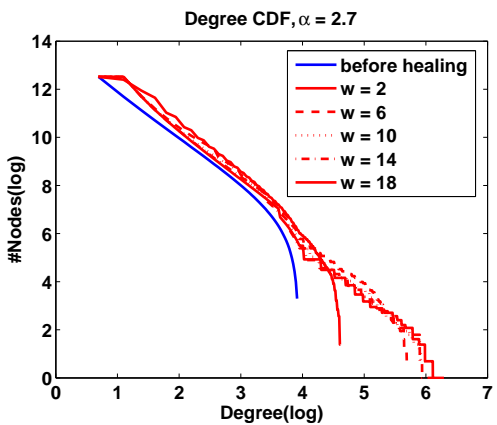}}}
     \hspace{0.2in}%
     \subfigure[]
       {%
        \centering
         \scalebox{0.5}{\includegraphics{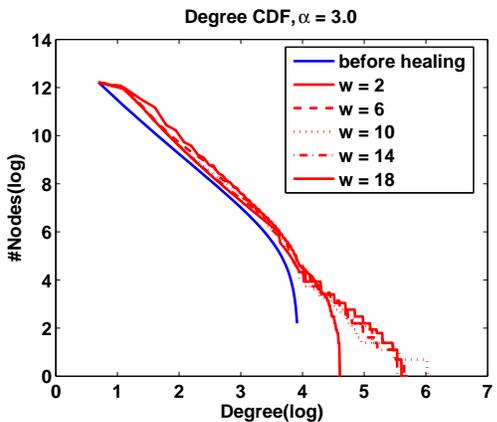}}}
\caption{Degree distribution before and after healing for various
values of power-law exponent $\alpha$ and stretch factors
$w$.}\label{fig:healing}
\end{figure}

We have performed the Monte Carlo simulations of the healing and
feedback algorithm. Fig. \ref{fig:healing} shows how the healing
algorithm works for different values of $\alpha$.

%

\section{Restoring Connectivity In A Fragmented Power-Law Network}
\label{sec:glueing} Any network will breakdown under a
sufficiently heavy targeted attack, and the question we ask is how many
random connections among the nodes in the fragmented network do we
need to establish before a giant connected component emerges. The
goal of this section is to show that both \emph{grown and random} power-law
networks, are very amenable to quick bootstrapping from an attack, and
with almost no global coordination. In particular,
we will show that even under very intense attacks, only an
infinitesimally small probability of success for new connections
is enough to quickly reconnect most of the broken network.

 For
example, consider a linearly preferentially grown PL network of
size $50,000$ and average degree $4$, that has undergone a very
heavy  targeted attack in which all nodes of degree more than $50$
are lost. For a particular simulation, the size of the largest
component was only 371 after the attack. Now assume that all the nodes try to
initiate only one random connection to some other node in the network in
a hope to restore the connectivity. If the probability of success
of each of these attempts is only 5\%, our simulations show that a
very large component of size more than 20,000 forms. This probability of success
will go asymptotically to zero as the network size increases. Of course,
other reconstruction algorithms will be required to repair the
topological damages to the network, as was the subject of the
algorithms in Section \ref{subsec:healing}.

The results in this section will follow the following pattern:
First we will argue that grown PL networks fragmented under a
targeted attack, have components whose size distributions are
heavy-tailed PL's. Thus, even though the average size of the
connected components is bounded, the variance is unbounded (or
very large). While a proof for grown networks is difficult in
general, we show analytically that for a linearly preferentially
grown \emph{tree} network (i.e., each node joining the network
makes exactly one preferential connection) the removal of the
highest degree node indeed creates disconnected components with a
PL size distribution. We then show, using the generating functions
formalism, that the same phenomenon can be observed in {\em random
PL networks} for special forms of targeted attacks.

Next, we prove that as long as we have components with
heavy-tailed size distribution, then only a vanishingly small
number of random connections will glue the fragments together into
a giant connected component.

\subsection{Distribution of the size of components after attack: Grown Power-law graphs}
Generally, instantaneous percolation (and hence attack) on grown
graphs is hard to analyze. For some special cases however, the
statistics of the connected components after a preferential attack
can be tracked. In this subsection, we compute the distribution of
the sizes of the connected components for graphs grown under the
preferential attachment (Barabasi-Albert graphs) with $m=1$: At
each time step, a new node is added to the network and initiates
\emph{one} preferentially targeted link. The probability that the $i^{th}$ node receives this link is $k(i,t)/\sum_{j=1}^t k(j,t)$, where $k(i,t)$ is the degree of the $i^{th}$ node at time $t$.

 At some time $t\gg 1$, a preferential attack occurs, and
deletes the oldest node $i=1$. As a result of such attack, the
network will be fragmented into many connected components. We now
show that the size of these components have a power-law
distribution, and in particular, the probability that a randomly
chosen component has size $C$ is $\propto C^{-3/2}$.

Before continuing, we need the following observations: (i) Since
$m=1$, the network topology is a tree. (ii) Take any node $i$ and
consider the subtree rooted at $i$ and consisting of all the nodes
younger than $i$. The size of this subtree, denoted by $T(i,t)$
can be calculated as follows: At time $t=i$, this subtree has had
only one node (with one link to some other node $j<i$), thus
$T(i,i)=1$, while there has been $i$ other links in the network.
The rate of change of $T(i,t)$ can be written as:
\[
\frac{\partial T(i,t)}{\partial t}=\frac{(2T(i,t)-1)}{2t}
\]
If $1$ can be neglected compare to $2T(i,t)$ (which is certainly
true when $t\gg i$), one gets: $T(i,t)=(t/i)$.

Now, note that from the rate-equations, the degree of a node
inserted at time $i$, at a later time $t$ is given by :
$k(i,t)=(t/i)^{\beta}$ for $\beta=1/2$. Thus the degree of the
first node is around $k(1,t)=t^{1/2}$.  Thus, once the first  node
is deleted, exactly $k(1,t)$ connected components will be created
(remember the network is tree). Lets enumerate these components by
the sets $C_1,C_2,...,C_{k(1,t)}$. Let $\kappa_j,j=1,2,...,k(1,t)$
be the oldest node in each $C_j$. By construction of the network,
the node $\kappa_j$ must have been connected to the first node
($i=1$). Now note that the size of the subtree $C_j$ is simply
$|C_j|=T(\kappa_j,t)=(t/\kappa_j)$. This will allow us to find the
probability that a randomly chosen connected component has size
$C$ as follows:
\begin{eqnarray*}
P_C&=&\frac{\#Components\;of\;size\; C}{k(1,t)}\\
&\propto&\left(\frac{Pr\{\kappa\;connected\;to\;1\}}{k(1,t)}
\times |\frac{\partial \kappa}{\partial
C}|\right)_{\kappa^*:T(\kappa^*,t)=C}\\
&\propto& \frac{(\kappa^*)^{\beta}}{\kappa^*}\times
|\frac{\partial \kappa}{\partial C}|_{\kappa^*:T(\kappa^*,t)=C}
\end{eqnarray*}
Now, note that $T(\kappa^*,t)=(t/\kappa^*)=C$, and $|\partial
\kappa/\partial C|=tC^{-2}$. Therefore:
\begin{eqnarray*}
P_C\propto \frac{C^{-\beta}}{C{-1}}\times C^{-2}=C^{-3/2}
\end{eqnarray*}

In \cite{N01}, Newman et. al, have shown that for any
\emph{static} random graph, the distribution of the size of the
components just before the phase transition and the appearance of
a giant connected component obeys the same scaling law as
$P_C\propto C^{-3/2}$. So interestingly, while the deletion of a
key node (the first node) fragments the network into many
components (around $t^{1/2}$ different pieces), the distribution
of the sizes of these components obeys a power-law. 

\begin{figure}[h]
\centering
    \scalebox{0.3}{\includegraphics{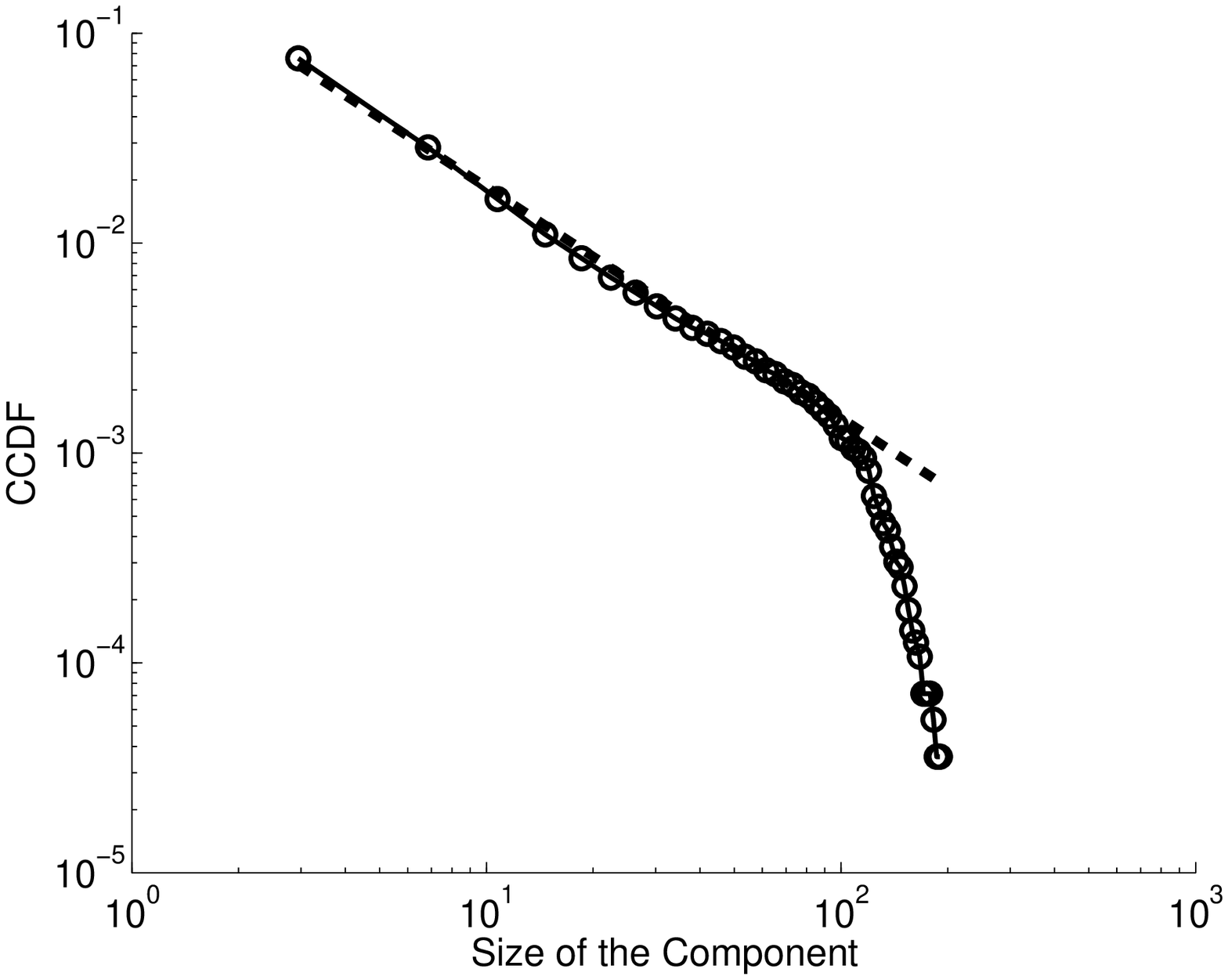}}
    \scalebox{0.3}{\includegraphics{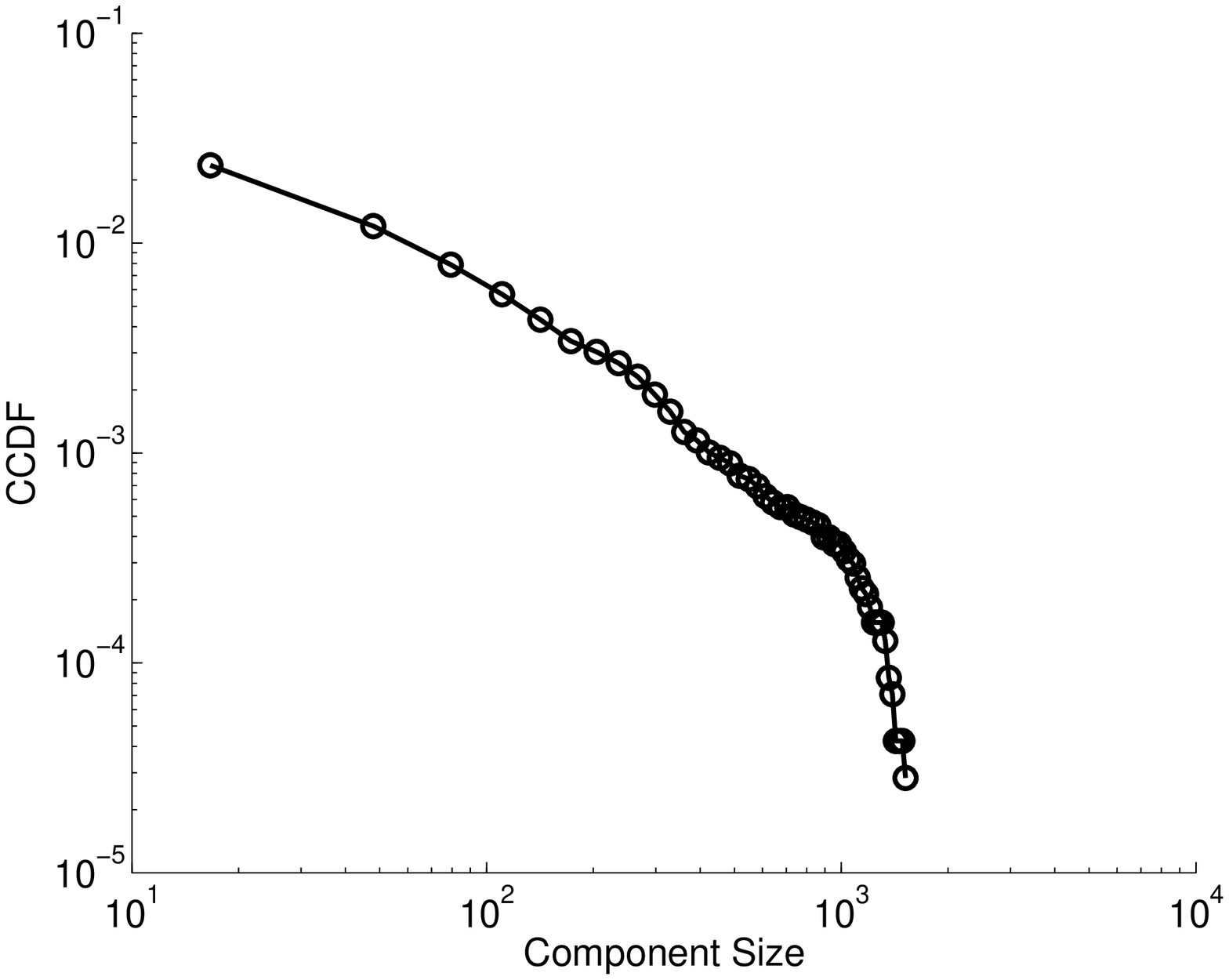}}
    \caption{ The distribution of the size of the connected components for heavily attacked networks of size 50,000 after a targeted attack that deleted all the nodes with degree greater than 50. (top)
    Linearly preferentially grown PL network (Barabasi-Albert model) when the average degree is 4. (bottom) Deletion-compensation networks of \cite{SR04}, with $\gamma \approx 2.3$ and average degree $6$. }
    \label{fig:comps}
\end{figure}

For more complex randomly grown graphs, no theory of percolation
currently exists to obtain the distribution of the connected
components. Simulations however indicate that the same
observations still hold for many such networks. One such
simulation is reported for networks grown with the deletion
compensation protocol introduced in \cite{SR04}, and is reported
in Fig. \ref{fig:comps}.

\subsection{Distribution of the size of components after attack: Static Power-law
graphs} We now show that the same observations hold for some form
of random targeted attacks on static power-law networks as well.
The size distribution of the connected components of any static
random network on a given degree distribution can be found
analytically using the generating functions formalism \cite{N01}.
In particular, the variance of the distribution of the size of
these components are derived analytically in Appendix \ref{ap1},
which also contains a brief introduction to generating functions
formalism.

Consider the generating functions of an attacked network:
\begin{eqnarray}
G_0(x)&=&\sum_{i=1}^{K} q_k p_k x^{-k}\\
G_1(x)&=&\frac{G_0'(x)}{G_0'(1)}\\
\end{eqnarray}
where $q_k$ is the probability that a node with degree $k$ is
deleted through the attack. This would define a form of targeted
attack if $q_k$ decreases with $k$. In particular, we examine a
targeted attack for which $q_k=b k^{-q}$ where $q$ is  a measure
of the how targeted the attack is and $b$ is a normalization
constant. Then, the generating functions of the attacked network
for an original power-law graph with exponent $\tau$ are:
\begin{eqnarray}
G_0(x)=\sum_{i=1}^{K} c_1 k^{-\tau-q} x^{-k}\\
\end{eqnarray}
\begin{equation}
G'_1(1)=c'\sum_{k=1}^K k^{-\tau-q+2}=O(K^{-\tau-q+3})
\end{equation}
for some positive constants $c,c'$ when $\tau+q\geq 2$.  In
particular, for the linear targeted attack , $q=1$, this value is
always finite for any $\tau > 2$.

It should be noted that such preferential attack would in effect
increase the value of $\tau$ by an amount of $q$. Thus following
the approach of Aiello et. al. \cite{Aiello00}, one can show that
no giant connected component will exist when:
$\tau+q>\beta_c\approx 3.478$.

Although the network might not have any giant connected component,
the variance of the distribution of the sizes might still diverge.
This is shown through $G''_1(1)$ (see Eqn. \ref{res}):
\[
G''_1(1)=\sum_{k=3}^K k(k-1)(k-2)b k^{-\tau-q}\propto
K^{-\tau-q+4}
\]
which diverges for any $q$ provided that $q<4-\tau$. In particular
for the linearly targeted attack, $q=1$, the variance of the size
of connected components diverges if $2<\tau<3$, even though the
average component size is finite at least when $\tau>2.349$.

\subsection{The Reverse Percolation Process}
\begin{figure}[h]
\centering
    \includegraphics[scale=0.6]{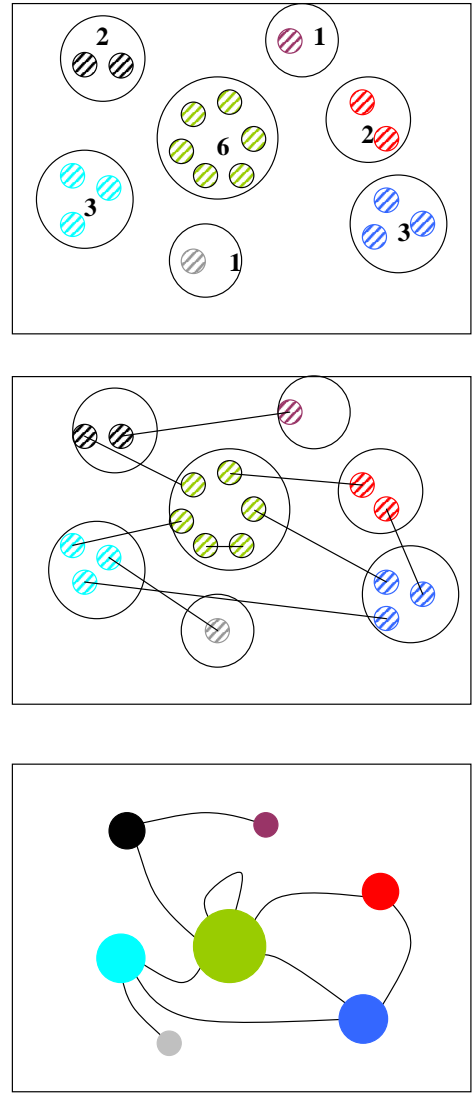}
    \caption{Constructing a random graph on degree distribution, $n_1=2,n_2=2,n_3=2,n_4=0,n_5=0,n_6=1$. For any node of degree $i$, $i$ dummy copies are created. Then a random matching is performed on this hyper graph.
    All the dummy nodes corresponding to one real node, are then collapsed into one nodes, to form the actual graph.}
    \label{fig:invperc}
\end{figure}

The main idea behind the results in this section is what we will refer to as the \emph{reverse percolation} process: Consider an attacked
network, as in Fig. \ref{fig:invperc}. Now lets start adding
random edges between the nodes of this network of many small components. As the
number of such random links increases, different components of the
network will start to \emph{glue} together until a giant connected
component occurs which contains most of the nodes of the network.
Lets call $Q(k)$ the distribution of the sizes of these small
components, that is, $Q(k)$ is the fraction of these components
that have size $k$. We claim that the reverse percolation process
corresponds to a percolation on a random graph with degree
distribution $P(k)\equiv Q(k)$. To see this we need to recall the
method with which a random graph with a prescribed degree
distribution $P(k)$ is built on $N$ nodes \cite{MR95} (see Fig.
\ref{fig:invperc}).

To construct a random graph with degree distribution $P(k)$, one
can proceed as follows \cite{moll95}: for any $k$, there will be
$NP(k)$ nodes of degree $k$ in the network, labelled as
$V^k_1,V^2_k,...,V^k_{NP(k)}$. For each of these nodes create $k$
dummy duplicate nodes. To be specific, call the duplicates of the
$i'th$ node as $V^k_{i,1},V^k_{i,2},...,V^k_{i,k}$. Now on this
hyper graph, one can start a random matching. After the matching,
one will collapse the duplicates of a node $i$ into one node, and
therefore all the links to the duplicates will now be links to the
collapsed node itself. A {\em reverse percolation process} with probability $p$
can be interpreted as introducing random edges, i.e., doing the random matching with a fraction $p$ of all the links in the hyper graph.

\begin{figure}[h]
\begin{center}
    \includegraphics[scale=0.6]{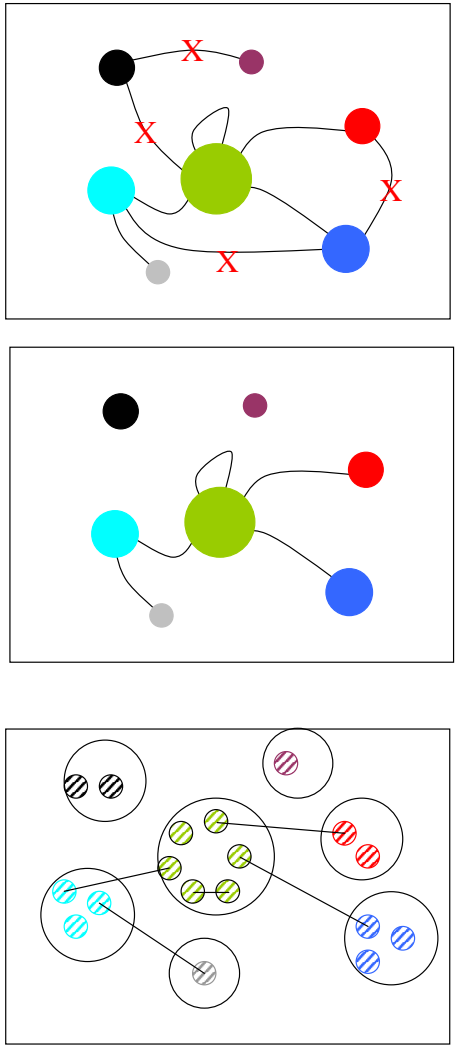}
    \caption{Percolation and Reverse Percolation Process: Bond percolation corresponds to random deletion of
    the links of a network. In the top figure, 4 out of 10 links of the random network generated in Fig. \ref{fig:invperc}, are deleted, resulting
    in  the middle graph. This was equivalent to doing the random graph matching on the hyper graph with only 10-4=6 links instead of
    10 links (the bottom figure). }
    \label{fig:invperc2}
\end{center}

\end{figure}

With this construction in mind, the validity of our claim is
readily understood. In the connectivity restoration process, {\em any
connected component can be viewed as the duplicate nodes of a
single node}, and the insertion of random edges can be viewed as a
(reverse) percolation process on a graph whose degree distribution
is equal to the distribution of the component sizes of the real
broken graph (see Fig. \ref{fig:invperc2}). In the usual percolation process,
you keep each edge with probability $p$, which in the random graph construction process, basically involves doing random matching with probability $p$, in the
network with duplicated nodes. With this
correspondence, many of the well known results for the percolation
process on random graphs with a given degree distribution can be
readily applied to the gluing process.

In particular, the percolation threshold, corresponding to the
probability of successful attempts required for a giant component
to appear can be calculated as: $q_c=\frac{\langle C
\rangle}{\langle C^2 \rangle-\langle C \rangle^2}$ where $\langle C
\rangle,\langle C^2 \rangle$ are the average and variance of the
distribution of the connected components (see for instance for
\cite{CNSW00}). While the average size of the connected components
in a heavily attacked network is finite, it is possible for the
variance of this distribution to be very large. In which case, the
corresponding percolation
 threshold will be small. In other words, one would only require
 an infinitesimally small fraction of random links to glue most of the
 disjoint components together. In particular, if the distribution
 of the size of the connected components follows a power-law
 distribution with exponent $2<\alpha<3$ and maximum component size
 $C_{max}\gg 1$, then only $O(C_{max}^{\alpha-3})$ successful random links per component is
 enough to ensure that most of the disconnected components are reconnected.

\section{Concluding Remarks}
We have addressed the issue of attack management in scale-free PL networks that are reactive, and can take local steps to combat attacks. In particular, we have shown how grown networks can cope with both random and linear-preferential progressive attacks, where nodes are deleted as the network grows.  We also presented a number of recovery  schemes, including repairing of sharp cutoffs in PL degree distributions, and restoration of connectivity in networks fragmented by large-scale targeted attacks. All these compensatory mechanisms are shown to be local, in the sense that global coordination among the nodes is not required, and the nodes initiate new edges only in reaction to changes in their immediate environment.

There are several interesting implications of the results presented in this paper in terms of complex network theory as well.  For example, Section II presents a network dynamic that leads to the emergence of PL degree distributions with exponential cutoffs; perhaps, such a mechanism can model existing networks where such degree distributions have been observed empirically. Similarly, when one studies the distribution of the size of the connected components in the networks generated by the dynamic in Section II, then one observes that there is always a giant connected component, but more interestingly, the rest of the components have a power-law size distribution.  Such a component size distribution has been observed, for example, in the world wide web (WWW) network, and one wonders if a low-grade preferential deletion of high-degree nodes in the web is one of the dynamical forces shaping the underlying connectivity structure.

\begin{appendix}

\section{Variance of the Distribution of the Size of Connected Components in Static Random Networks}\label{ap1}
 Consider a
random network on a given degree distribution $P(k)$ \cite{N01}, that is the
probability that a randomly chosen node has degree $k$ is ,
$P(k)$.
 Define
\begin{eqnarray*}
G_0(x)&=&\sum_{i=1}^{K} P(k) x^{-k}\\
G_1(x)&=&\frac{G_0'(x)}{G_0'(1)}\\
\end{eqnarray*}
as the generating functions of the degree of a node arrived at by
choosing a random node or link respectively.

Similarly, one can define $H_1(x),H_0(x)$ as the generating
function of the size of the connected components arrived at by
following a random link and node respectively.

When a giant connected component does not exist and therefore the
graph is tree like, these four functions should be related through
the following consistency equation:
\begin{eqnarray}\label{s1}
H_1(x)=xG_1(H_1(x))\\
 H_0(x)=xG_0(H_1(x))
\end{eqnarray}
The average distribution of the sizes of the connected components
are:
\[
\ave{s}=H_1'(1)=1+\frac{G_1'(1)}{1-G_1'(1)}
\]
The phase transition happens when $G_1'(1)=1$ and a giant
connected component appears (or disappears).

The known literature has always been interested in the average
size of the connected components due to its phase transition and
physically measurable properties. We will however examine the
properties of the second moment of the distribution of the size of
the connected components.

Twice differentiating $H_1(x)$ in (\ref{s1}):
\begin{eqnarray}\label{s2}
H''_1(x)&=&G'_1(H_1(x))H'_1(x)+H'_1(x)G'_1(H_1(x))\\
&+&xH''_1G'_1(H_1(x))+x(H'_1(x))^2G''_1(H_1(x))\nonumber
\end{eqnarray}
Once differentiating $H_1(x)$ in (\ref{s1}) results in:
\[
H'_1(1)=\frac{2}{1-G'_1(1)}
\]
Plugging this into (\ref{s2}) evaluated at $x=1$ will result in:
\begin{equation}\label{res}
E\{size^2\}=H''_1(1)=\frac{4G'_1(1)}{(1-G'_1(1))^2}+\frac{4G''_1(1)}{(1-G'_1(1))^3}
\end{equation}

 This second moment can diverge in two cases. First,
if $G'_1(1)\rightarrow 1$, and second $G''_1(1)\rightarrow
\infty$. When a giant connected component does not exist,
$G'_1(1)<1$, the second moment can still diverge provided that
$G''_1(1)$ diverges.

\end{appendix}
\bibliography{final_attack}
\end{document}